\documentclass[prd,11pt,aps,epsfig,showpacs,nofootinbib,floatfix]{revtex4}
\usepackage{epsfig,amssymb,amsmath,graphics, amssymb}
\usepackage{textcomp}
\usepackage{gensymb}
\usepackage{hyperref}

\newcommand{\ps}{$e^{+}$}
\newcommand{\el}{$e^{-}$}

\newcommand\beq{\begin{equation}}
\newcommand\eeq{\end{equation}}
\newcommand\bea{\begin{eqnarray}}
\newcommand\eea{\end{eqnarray}}
\newcommand\bi{\begin{itemize}}
\newcommand\ei{\end{itemize}}
\newcommand\ben{\begin{enumerate}}
\newcommand\een{\end{enumerate}}
\newif\ifboo \boofalse

\def\lsim{\mathrel{\rlap{\lower4pt\hbox{\hskip1pt$\sim$}}
    \raise1pt\hbox{$<$}}}         
\def\gsim{\mathrel{\rlap{\lower4pt\hbox{\hskip1pt$\sim$}}
    \raise1pt\hbox{$>$}}}         

\begin{document}

\textheight=22.8cm



\title {Results from PAMELA, ATIC and FERMI : Pulsars or Dark Matter ?}
\date{\today}

\author{Debtosh Chowdhury}
\email{debtosh@cts.iisc.ernet.in}
\affiliation{Centre for High Energy Physics, Indian Institute of Science, Bangalore, India}
\author{Chanda J. Jog}
\email{cjjog@physics.iisc.ernet.in}
\affiliation{Department of Physics, Indian Institute of Science, Bangalore, India}
\author{Sudhir K. Vempati}
\email{vempati@cts.iisc.ernet.in}
\affiliation{Centre for High Energy Physics, Indian Institute of Science, Bangalore, India}

\begin{abstract}
It is well known that the dark matter dominates the dynamics of galaxies and
clusters of galaxies. Its constituents remain a  mystery despite an assiduous search for them over
the past three decades. Recent results from the satellite-based  PAMELA experiment detect an excess in the positron
fraction at energies between $10-100$ GeV in the secondary cosmic ray spectrum. Other experiments  namely ATIC, HESS and
FERMI  show an excess in the total electron  (\ps + \el) spectrum for energies greater 100 GeV. These excesses in the positron
fraction as well as the  electron spectrum could arise in local astrophysical processes like pulsars, or can be attributed to the annihilation
of the dark matter particles.  The second possibility gives clues to the possible candidates for the dark matter in galaxies
and other astrophysical systems. In this article, we give a  report of these exciting developments.
\end{abstract}
%
\maketitle
\vskip .6 true cm
\section{Introduction}
The evidence for the existence of dark matter in various astrophysical systems
has been gathering over the past three decades. It is now well-recognized that the presence of dark matter is required in order to explain
the observations of galaxies and other astrophysical systems on larger scales. The clearest support for the existence of dark matter
comes from the now well-known observation of nearly flat rotation curves or constant rotation velocity in the outer parts of
galaxies \cite{rubin,Sofue:2000jx}. Surprisingly the rotation velocity is observed to remain nearly constant till the last
point at which it can be measured\footnote{In the absence of dark matter, one would expect that the curves to fall off as we move
towards the outer parts of the galaxy.}. The simple principle of rotational equilibrium then tells one that the
amount of dark to visible mass must increase at larger radii. Thus the existence of the dark matter is deduced
from its dynamical effect on the visible matter, namely the stars and the interstellar gas in galaxies.

The presence of dark matter in the elliptical galaxies is more problematic to ascertain since
these do not contain much interstellar hydrogen gas which could be used as a tracer of their dynamics, and
also because these galaxies are not rotationally supported. These galaxies are instead supported
by pressure or random motion of stars (see Binney \cite{Binney:1987} for details of physical properties of the
spiral and elliptical galaxies). As a result, the total mass cannot be deduced using the rotation curve for elliptical galaxies. 
Instead, here the motions of planetary nebulae which arise from  old, evolved stars, as well as lensing, have been used to trace the dark matter \cite{dekel05}.
 The fraction of dark matter at four effective radii is still uncertain with values ranging from 20\% to 60\% given in the
literature, for the extensively studied elliptical galaxy NGC 3379 \cite{mamon}.

Historically the first evidence for  the unseen or dark matter was found in \textit{clusters} of galaxies.
Assuming the cluster to be in a virial equilibrium, the total or the virial mass can be deduced from the observed kinematics.
Zwicky \cite{zwicky} noted that there is a discrepancy of a factor of $\sim$ 10 between the observed mass in clusters of galaxies
and the virial mass deduced from the kinematics. In other words, the random motions are too large for the cluster to be bound
and a substantial amount of dark matter ($\sim$ 10 times the visible matter in galaxies) is needed for the clusters of galaxies
to remain bound. This discrepancy  remained a puzzle for over four decades, and was only realized to be a part of the general
trend after the galactic-scale dark matter was discovered in the late 1970's.

On the much larger cosmological scale, there has been some evidence for non-baryonic dark matter from theoretical estimates of 
primordial elements during  Big Bang Nucleosynthesis and measurements of them, particularly, primordial deuterium. Accurate
measurements of the Cosmic Microwave Background Radiation (CMBR) could as well give information about the total dark matter relic density of the 
Universe. The satellite based COBE experiment was one of the first experiments to provide accurate `` mapping" of the CMBR\cite{cobe}. 
The recent high precision determination of the cosmological parameters using Type I supernova data \cite{sndata} as well as 
precise measurements of the cosmic background radiation by the WMAP collaboration \cite{wmap,kom08} has 
pinpointed the total relic dark matter density in the early universe with an accuracy of a few percent. Accordingly, dark matter
 forms almost 26\% of all the matter density of the universe, with visible matter about 4\% and the dark energy roughly about 70\% 
 of the total energy density. This goes under the name of $\Lambda$CDM model with $\Lambda$ standing for dark energy and denoted
  by the Einstein's constant, and CDM standing for Cold Dark Matter \cite{cosmoreviews}.

 Numerical simulations for the currently popular scenario of galaxy formation, based on the ${\Lambda}$CDM model, 
predicts a universal profile for the dark matter in halos of spherical  galaxies \cite{NFW97}. While this model was initially successful, over the years many
discrepancies between the predictions from it and the observations have been pointed out. The strongest one has been the
`cusp-core' issue of the central mass distribution. While Navarro {\it et al.}  \cite{NFW97} predict a cuspy\footnote{Sharp increase in the density 
at the centre.}  central mass distribution,
the observations of rotation curves of  central regions of galaxies, especially the low surface brightness galaxies, when modeled
show a flat or cored density distribution \cite{BMBR01}.

A  significantly different alternative to the dark matter, which can be used  to explain the rotation curves of the galaxies and clusters  was proposed 
early on by Milgrom. He claimed that \cite{Milgrom:1983ca} for low accelerations, Newtonian law has to be modified by addition of a small repulsive term. 
This idea is known as \lq{}MOND\rq{} or the MOdified Newtonian Dynamics. While initially this idea was not taken seriously by the majority of astrophysics community, it has gained more acceptance in the recent years. For example some of the standard features seen in galaxies such as the frequency of bars can be better explained under the MOND paradigm, see Tiret \textit{et al.} \cite{Tiret:2007dd}. For a summary of the predictions and comparisons of these two alternatives (dark matter and MOND), see Combes \textit{et al.} \cite{Combes:2009ab}.

So far the most direct empirical proof for the existence of dark matter, and hence the evidence against MOND, comes from the study of the so-called Bullet cluster\cite{Clowe:2006eq}. This is a pair of galaxies undergoing a supersonic collision at a redshift of $\sim 0.3$. The main visible baryonic component in clusters is hot, X-ray emitting gas. In a supersonic collision, this hot gas would collide and be left at the center of mass of the colliding system while the stars will just pass through since they occupy a small volume\footnote{This is exactly analogous to the reason why the atomic hydrogen gas from two colliding galaxies is left at the center of mass while the stars and the molecular gas pass through each other unaffected, as proposed and studied by Valluri \textit{et al.}\cite{jog:1990} to explain the observed HI deficiency but normal molecular gas content of galaxies in clusters.}.

In the Bullet cluster, 
the gravitational potential as traced by the weak-lensing shows peaks that are separated from the central region traced by the hot gas. In MOND, these 
two would be expected to coincide\footnote{The relativistic MOND theory \cite{Bek1} proposed by Bekenstein  could be used to explain the Bullet Cluster \cite{Bek2}.},
 since the gravitational potential would trace the dominant visible component namely the hot gas, while if there is dark matter it would be expected to peak at the
location of the stellar component in the galaxies. The latter case is what  has been observed as can been seen in  Fig.1 of \cite{Clowe:2006eq}. 
For the rest of the article, we will not  consider the MOND explanation, but instead take the view point that the flat rotation curves of galaxies and 
clusters at large radii as an evidence for the 
existence of dark matter. Furthermore, we believe that the dark matter explanation is much simpler and more natural compared to the MOND explanation. 

Despite the fact that the existence of dark matter has been postulated  for over three decades, there is still no consensus
of what its constituents are. This has been summarized well in many review articles. Refs \cite{trimble87, AKS09} are couple
of examples  that span from the early to recent times on this topic.  Over the years, both astrophysicists as well as particle physicists have speculated
on the nature of dark matter. 

The baryonic dark matter in the form of low-mass stars, binary stars, or Jupiter-like massive planets were ruled out early on
(see \cite{trimble87} for a summary).  From  the amount of dark matter required to explain the flat rotation curves, it can be shown
that  the number
densities  required of these possible constituents would be  large, and hence it would be hard to hide these massive objects. 
Because, if present in these forms,
they should have been detected either from their absorption  or from their emission signals. It has also been proposed that the
galactic dark matter could be in the form of dense, cold molecular clumps \cite{PCM94}, though this has not yet been
detected. This alternative cannot be expected to explain the dark matter necessary to
``fit"  the observations of clusters, or indeed the elliptical galaxies since the latter have very little interstellar gas.

There is also a more interesting possibility of the dark matter being essentially of baryonic nature, but due to the 
dynamics of the QCD phase transition in the early universe which left behind a form of  \textit{cold} quark-gluon-plasma,
 the baryon number content of the dark matter is hidden from us. This idea was first proposed by Witten in 1984 \cite{witten}, who
called these quantities as quark nuggets. An upper limit on the total number of baryons in a quark nugget  is determined by
the baryon to photon ratio in the early universe (See for example \cite{raha1}).  Taking in to consideration these constraints, 
it is possible to fit the observed relic density with a mass (density) distribution of the quark nuggets \cite{raha2}.  
For the observational possibilities of such quark  nuggets, see for example,  Ref.\cite{nuggetstudy}. 

From a more fundamental point of view, it is not clear what kind of elementary particle could form dark matter.
The standard model of particle physics describes all matter to be made up of quarks and leptons of which
neutrinos are the \textit{only} ones which can play the role of dark matter as they are electrically
neutral. However  with the present indications from various neutrino oscillation experiments putting the standard model 
 neutrino masses  in the range  $\lesssim 1$ eV \cite{valleneutrinoreview} 
they will not form significant amount of dark matter. There could however, be non-standard sterile neutrinos with masses  of the order of keV-MeV  which
could form \textit{warm}\footnote{Depending on the mass of the particle which sets its thermal and relativistic properties,  dark matter can be classified 
as \textit{ hot}, \textit{warm} and \textit{cold} \cite{Peacock:1999ye}.}dark matter (for reviews, see Refs. \cite{strumiavissani,julien1}). 
 Cold Dark Matter (CDM), on the other hand, is favored over the {\it warm} dark matter 
by the hierarchical clustering observed in numerical simulations for large scale structure formation, see for example Ref.\cite{Peacock:1999ye}. Recent analysis including X-ray flux observations
from Coma Cluster and Andromeda galaxy have shown that the room for sterile neutrino warm dark matter is highly constrained \cite{julien2}.
However, if one does not insist that the total relic dark matter  density is due to sterile neutrinos then, it is still possible that they form 
a sub-dominant \textit{warm} component of the total dark matter \cite{silk2} relic 
density\footnote{On the other hand, if the neutrinos are not thermally produced and their production is suppressed 
like in models with low reheating temperature \cite{gelmini1}, it is possible to weaken the cosmological bounds, especially from 
extra galactic radiation and distortion of CMBR spectra \cite{gelmini2}. See also \cite{julupdate}.}.

The Standard Model thus,  needs to be extended to incorporate a dark matter candidate. The simplest
extensions would be to just include a new particle which is a singlet under the SM gauge group (\textit{i.e.},
does not carry  the Standard Model interactions). Further, we might have to impose an additional symmetry
under which the Dark Matter particle transforms non-trivially to keep it stable or at least sufficiently long lived 
with a life time typically larger than the age of the universe.  Some of the simplest models would just
involve adding additional light ($\sim$ GeV) scalar particles to the SM and with an additional $U(1)$ symmetry 
(see for example, Boehm \textit{et al.} \cite{fayet1}).  Similar extensions of SM can be constructed
with fermions too \cite{fayet2,wells}. An interesting aspect of these set of models is that they can be tested
at existing $e^+ e^-$ colliders like for the example, the one at  present at Frascati, Italy \cite{dreeschoudhury}.
  A heavier set of dark matter candidates can be achived by extending
the Higgs sector by adding additional Higgs scalar doublets. These go by  the name of \textit{inert} Higgs models \cite{Barbieri:2005kf,marajaji}. In this extension, there is a \textit{additional} neutral  higgs boson which does not have SM gauge interactions (hence inert), which can be a dark matter candidate. With the inclusion of this extra inert higgs doublet, the SM particle spectrum has some added features, like it can evade the \lq\lq{}{\it naturalness problem}\rq\rq{} up to 1.5 TeV  while preserving the perturbativity of Higgs couplings up to high scales and  further it is consistent with the electroweak precision tests \cite{Barbieri:2006dq}.

On the other hand, there exist extensions of the Standard Model (generally labeled Beyond Standard Model (BSM) 
physics) which have been constructed to address a completely
different problem called the hierarchy problem. The hierarchy problem addresses the lack the symmetry 
for the mass of the Higgs boson in the Standard Model and the consequences of this in the light of the large
difference of energy scales between the weak interaction  scale ($\sim 10^2 $ GeV)  and the quantum gravity or 
grand unification scale ($\sim 10^{16} $ GeV).  Such a huge difference in the energy scales could destabilize the Higgs mass due to quantum corrections.
To protect the Higgs mass from these dangerous radiative corrections, new theories such as supersymmetry, large extra dimensions and little Higgs have
been proposed.   It turns out that  most of these BSM physics models  contain a particle which can be the dark matter. 
 A few examples of these theories and the corresponding candidates for dark matter are as follows. (i) Axions are pseudo-scalar particles which appear in theories with Peccei-Quinn symmetry \cite{Peccei:1977hh,Peccei:2006as} proposed  as  solution  to the \textit{strong CP} problem of the standard model.  They also appear in Superstring theories which are theories of quantum gravity.  The present limits on axions are \cite{Bertone:2004pz} extremely strong from astrophysical data. 
 In spite of this, there is still room for axions to form a significant part of the dark matter relic density. 

(ii) Supersymmetric theories \cite{Martin:1997ns,Drees:2004jm} which incorporate fermion-boson interchange symmetry are proposed as extensions  of Standard Model to protect the Higgs mass from large radiative corrections. The dark matter candidate is the lightest supersymmetric particle (LSP) which is stable or sufficiently long lived
as mentioned before\footnote{The corresponding symmetry here is called $R$-parity. If this symmetry is exact, the particle is stable. If is broken very  mildly, the LSP
could be  sufficiently long lived, close to the age of the universe.}. Depending on how supersymmetry is broken \cite{jun96}, there are several possible dark matter candidates in these models. In some models, the lightest supersymmetric particle and hence the dark matter candidate is a neutralino. The neutralino is a linear combination of super-partners  of $Z, \gamma$ as well as the neutral Higgs bosons\footnote{The neutralino could be either gaugino dominated or higgsino dominated depending on the composition. It turns out that neutralino composition should be sufficiently \textit{ well-tempered}\cite{ArkaniHamed:2006mb} to explain the observed
relic density. While one might debate the some what philosophical requirement of `fine-tuning', it is now known that in simplest models of supersymmetry breaking, like mSUGRA, only special regions in the parameter space, corresponding to the special conditions in the neutralino-neutralino annihilation channels satisfy 
 the relic density constraint \cite{dreesdjouadi}. }. The other possible candidates are the super-partner of the graviton, called the gravitino and the super-partners
 of the axinos, the scalar saxion and the fermionic axino. These particles also can explain the observed relic density \cite{covireview}. 

(iii) Other classic extensions of the Standard Model either based on additional space dimensions or larger symmetries also have dark matter candidates. 
 In both versions of the extra dimensional models, \textit{i.e.,} the Arkani-Hamed, Dimopoulos, Dvali (ADD) \cite{ArkaniHamed:1998rs, ArkaniHamed:1998nn} and Randall-Sundrum (RS) \cite{Randall:1999ee, Randall:1999vf}, models, the lightest Kaluza-Klein particle\footnote{The extra space dimensions are compactified. The compact
 extra dimension manifests it selves in ordinary four dimensional space-time as an infinite tower of massive particles called Kaluza-Klein (KK) particles.} can be considered as the dark matter candidate \cite{Servant:2002aq,Hooper:2007qk, che02,Bertone:2002ms}. Similarly, in the little-Higgs models where the Higgs boson is a pseudo-Goldstone boson of a much larger symmetry, a symmetry called T-parity \cite{Hubisz:2004ft} assures us a  stable and neutral particle which can form the dark matter. 
Very heavy neutrinos with masses of $\mathcal{O}(100~\text{GeV} - 1~\text{TeV})$ can also naturally appear within some classes of Randall-Sundrum and 
Little Higgs models.  Under suitable conditions, these neutrinos can act like cold dark matter. (For a recent study, please see \cite{geraldine} ).  
In addition to these  particles, more exotic candidates like simpzillas \cite{simpzillas} and wimpzillas \cite{kolb2} 
with masses close to the GUT scale ($\sim 10^{15}$ GeV) have also been proposed in the literature. Indirect searches like 
ICECUBE \cite{icecube} (discussed below) already have strong constraints on simpzillas.

\section{Dark Matter Experiments }
If the dark matter candidate is indeed a new particle and it has interactions other than gravitational
interactions\footnote{It cannot have electromagnetic interactions as this would mean it is charged, and it cannot have strong interactions as this
would most likely mean it would be baryonic in form - both these prospects are already ruled out
by experiments.}, then the most probable interactions it could have are the weak interactions\footnote{In spite of being electrically neutral, dark-matter particle can have a nonzero electric and/or magnetic dipole moment, if it has a nonzero spin. In such a case the strongest constraint comes from Big Bang Nucleosynthesis. Interested readers are referred to the paper by Kamoinkowski {\it et al.} \cite{Sigurdson:2004zp} and particularly Fig.~1 therein.}.
 This weakly interacting particle, dubbed as WIMP (Weakly Interacting Massive Particle)
could interact with ordinary matter and leave traces of its nature.  There are two ways in which the WIMP
could be detected (a) Direct Detection:  here one looks for the interaction of the WIMP on a target, the target being
typically nuclei in a scintillator. It is expected that the  WIMPs  present all over the  galaxy  scatter off  the target nuclei once in a while. 
 Measuring the  recoil of the nuclei in these rarely occurring events   would give us information about the  properties of the WIMP.  The scattering cross section
would depend on whether it was elastic or inelastic  and is a function of  the spin of the WIMP \footnote{More generally, 
the WIMP-Nucleon cross section can be divided as (i) elastic spin-dependent (eSD), (ii) elastic spin-independent (eSI) , 
(iii) in-elastic spin-dependent ( iSD) and (iv) in-elastic spin-independent (iSI).}.   There are more
than 20 experiments  located all over the world, which are currently looking for WIMP through this technique.
Some of them  are DAMA, CDMS, CRESST, CUORICINO, DRIFT {\it etc.} (b) Indirect detection :  when WIMPs cluster together in the galatic halo, 
they can annihilate with themselves giving rise to electron-positron pairs,  gamma rays, proton-anti-proton pairs, neutrinos {\it etc.} The flux of such radiation is directly proportional to the annihilation rate and the the WIMP matter density.  Observation of this radiation could lead to information about the
mass and the cross section strength of the WIMPs.  Currently, there are several experiments which
are looking for this radiation\footnote{These are typically the same experiments which measure the cosmic ray spectrum.
For a comprehensive list of all these experiments and other useful information like propagation packages, please have 
a look at: \\ \url{http://www.mpi-hd.mpg.de/hfm/CosmicRay/CosmicRaySites.html}.}  (i) MAGIC, HESS, CANGAROO, FERMI/GLAST, 
EGRET {\it etc.} look for the gamma ray photons. (ii) HEAT, CAPRICE, BESS, PAMELA, AMS  can observe anti-protons and positron flux.
(iii)Very highly energetic neutrinos/cosmic rays $\sim $ a few TeV to multi-TeV  can be  observed by large detectors like
 AMANDA, ANTARES, ICECUBE {\it etc.} (for a more detailed discussion see \cite{Bertone:2004pz,pijush1}).

Over the years, there have been indications of  presence of the dark matter through both direct and indirect experiments.
The most popular of  these signals are INTEGRAL and DAMA results (for a nice discussion on these topics 
please see, \cite{Hooper:2009zm}). INTEGRAL (International Gamma -Ray Astrophysics Laboratory) is a satellite 
based experiment looking
for gamma rays in outer space. In 2003, it has observed a very bright emission of the 511 keV photons 
from the Galactic Bulge \cite{integral1} at the centre. The 511 KeV line is special as  it is dominated by 
$e^+ e^-$ annihilations via the positronium. The observed rate of (3-15) $\times 10^{42}$ positrons/sec  
in the inner galaxy was much larger than the expected rate from pair creation via cosmic ray interactions
with the interstellar medium in the galactic bulge by orders of magnitude\footnote{It should be noted that the
Integral spectrometer has a very good resolution of about 2 KeV over a range of energies 20 keV to 8 MeV.}.  
Further, the signal is approximately spherically symmetric with very little positrons from galactic bulge 
contributing to the signal \cite{integral2}. Several explanations have been put forward to explain this excess. 
Astrophysical entities like hypernovae, gamma ray burts and X-ray binaries
have been proposed as the likely objects contributing to this excess. On the other hand, this signal can also 
be attributed to the presence of dark matter which could annihilate itself giving rise to electron-positron pairs. 
To explain the INTEGRAL signal in terms of dark matter, extensions of Standard Model involving  light $\sim (\text{MeV} -\text{GeV})$ 
particles and light gauge bosons ($\sim \text{GeV}$) are ideally suited. These models which have been already
reviewed  in the previous section, can be probed directly at the existing and future $e^+ e^-$ colliders and hence could be tested. 
Until further confirmation from either future astrophysical experiments or through ground based colliders comes about, 
the INTEGRAL remains an `anomaly' as of now. 

While the INTEGRAL is an indirect detection experiment, the DAMA (DArk MAtter ) is a direct detection experiment
located in the Gran Sasso mountains of Italy. The target material consists of highly radio pure NaI crystal scintillators;
the scintillating light from WIMP-Nucleon scattering and recoil is measured.  The experiment looks for an annual 
modulation of the signal as the earth revolves around the sun \cite{dama1}. Such modulation of the signal is due
to the gravitational effects of the Sun as well as rotatory motion of the earth\footnote{Looking for such modulations 
further limit any systematics present in the experiment.}. DAMA and its upgraded version DAMA/LIBRA have collected
data for seven annual cycles and four annual cycles respectively\footnote{These results have been recently updated
with six annual cycles for DAMA/LIBRA; the CL has now moved up to $8.9 \sigma$ \cite{dama2}.}. Together they have reported an annual modulation  
 at 8.2$\sigma $ confidence level. If confirmed, the DAMA results would be the first direct experimental evidence for
 the existence of WIMP dark matter particle. However, the DAMA results became controversial as this positive signal
has not been confirmed by other experiments like XENON and CDMS, which have all reported null results in the
 spin independent WIMP-Nucleon scattering signal region. 
 
 The Xenon 10 detector also at Gran Sasso laboratories uses a Xenon target while measuring simultaneously the 
 scintillation and ionization produced by the scattering of the dark matter particle. The simultaneous measurement
 reduces the background significantly down to $4.5~ \text{KeV}$.  With a fiducial mass of 5.4 Kg, they set an upper
 limit of WIMP-Nucleon spin independent cross section to be 8.8 $ \times ~10^{-44} \text{cm}^2$ for a WIMP mass of 100 GeV\cite{xenon1}. 
An upgraded version Xenon 100 has roughly double the fiducial mass has started taking data from Oct 2009. In
the first results, they present null results, with upper limits of about $3.4 \times 10^{-44} \text{cm}^2$ for 55 GeV WIMPs \cite{xenon2}.
These results severely constraint interpretation of the DAMA results in terms of  an elastic spin independent 
WIMP-nucleon scattering. 

The CDMS (Cryogenic Dark Matter Search ) experiment has 19 Germanium detectors located in the underground
Soudan Mine, USA. It is maintained at temperatures $\sim 40 \text{mK}$ (milli-Kelvin). Nuclear recoils can be 
``seen" by measuring the ionisation energy in the detector.  Efficient separation between electron recoils and 
nuclear recoils is possible by employing various techniques like signal timing and measuring the ratios of the
ionization energies. Similar to Xenon, this experiment \cite{cdms1} too reported null results in the signal region\footnote{The
final results have a non-zero probability of two events in the signal region, we comment on it in the next section.}
and puts an upper limit $\sim 4.6 \times 10^{-44} \text{cm}^{2} $  on the WIMP-Nucleon cross-section 
for a WIMP mass of around 60 GeV. 

 The CoGeNT (Cryogenic Germanium Neutrino Technology) collaboration runs another recent experiment 
 which uses ultra low noise Germanium detectors. It is  also located in the Soudan Man, USA. The experiment
 has one of the lowest backgrounds below 3 KeVee ( KeV electron equivalent (ee) ionisation energy). It could
 further go down  to 0.4 KeVee, the electron noise threshold. The first initial runs have again reported null
 results \cite{cogent1} consistent with the observed background. At this point, the experiment did not have the
 sensitivity to confirm/rule out the DAMA results. However, later runs have shown some excess events over the
 expected background in the low energy regions \cite{cogent2}. While, the collaboration could not find a suitable explanation
 for this excess ( as of now) there is a possibility of  these \textit{excess} events having their origins in a very 
 light WIMP dark matter particle. However, care should be taken before proceeding with this interpretation as 
 the CoGeNT collaboration does not distinguish between electron recoils and nucleon recoils\cite{weinercogent}. 

In the light of these experimental results, the DAMA results are hard to explain. One of the ways out to make
the DAMA results consistent with other experiments is to include an effect called ``channelling" which could be
present only in the NaI crystals which DAMA uses. However, even the inclusion of this effect does not improve
the situation significantly. To summarize, the situation is as follows for various interpretations of the WIMP-Nucleon
cross section. For eSI (elastic Spin Independent) interpretation, the DAMA regions are excluded by both 
CDMS as well as Xenon 10.  This is irrespective of whether one considers the channeling effect or not. It is also 
hard to reconcile DAMA results with CoGeNT in this case. For  elastic Spin Dependent (eSD) interpretation, the
DAMA and CoGeNT results though consistent with each other are in conflict with other experiments. 
For an interpretation in terms of WIMP-proton scattering, the results are in conflict with several experiments 
like SIMPLE , PICASSO etc. On the other hand, an interpretation in terms of WIMP-neutron scattering is ruled
out by XENON and CDMS data.  For the inelastic dark matter interpretations, spin -independent cross section 
with a medium mass  ($\sim 50 ~\text{GeV}$) WIMP is disfavored by CRESST as well as CDMS data. For a low
mass (close to 10 GeV) WIMP, with the help of channeling in the NaI crystals, it is possible to explain the 
DAMA results, in terms of spin- independent inelastic dark matter - nucleon scattering. However, the relevant
parameters (dark matter mass and mass splittings) should be fine tuned and further, the WIMP velocity 
distribution in the galaxy should be close to the escape velocity.  Inelastic Spin dependent interpretation
of the DAMA results is a possibility (because it can change relative signals at different experiments 
\cite{koppzupan} ) which does not have significant constraints from other experiments. However, it has been 
shown\cite{weinercogent} that  inelastic dark matter either with spin dependent or spin independent interpretation of the DAMA
 results is difficult to reconcile with the CoGeNT results,  unless one introduces substantial exponential 
 background in the CoGeNT data.

\section{The Data} 

The focus of the present topical review is a set of new experimental results which have appeared over the
past year. In terms of the discussion in the previous section, these experiments follow ``indirect" 
methods to detect dark matter.  The data  from these experiments seems to be pointing to 
either ``discovery" of the dark matter  or some yet non-understood  new astrophysics  being operative within the vicinity of our Galaxy. 
 The four main experiments which have led to this excitement are (i) PAMELA\cite{Adriani:2008zr} 
 (ii) ATIC\cite{:2008zzr} (iii) HESS\cite{Collaboration:2008aaa} and
 (iv) FERMI\cite{Abdo:2009zk}. All of these experiments involve international collaborations spanning several 
 nations. While PAMELA and FERMI are satellite based experiments,
ATIC is a balloon borne experiment and HESS is a ground based telescope.  All these experiments contain 
significant improvements in technology over previous generation experiments of similar type.  The H.E.S.S 
experiment  has a factor $\sim 10$ improvement in $\gamma$-ray flux sensitivity  over previous experiments 
largely due to its superior  rejection of the hadronic background. Similarly, ATIC is the next generation balloon 
based experiment equipped to have higher resolution as well as larger statistics.  Similar statements also hold
for the satellite based experiments, PAMELA and FERMI. It should be noted that the satellite based experiments
have some inherent advantages over the balloon based ones.  Firstly, they have enhanced data taking period,
unlike the balloon based ones  which can take data only for small periods. And  furthermore, these experiments 
also do not have problems with the  residual atmosphere  on the top of the instrument  which plagues the balloon 
based experiments.

\begin{figure}[ht]
\includegraphics[width=0.40\textwidth]{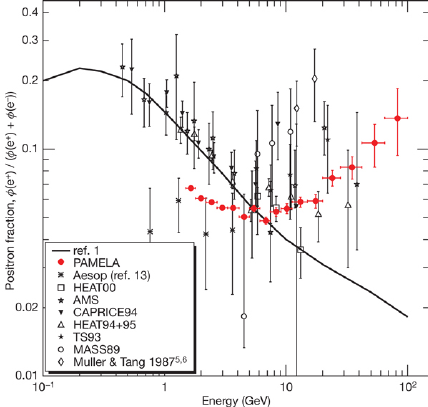}
\includegraphics[width=0.50\textwidth]{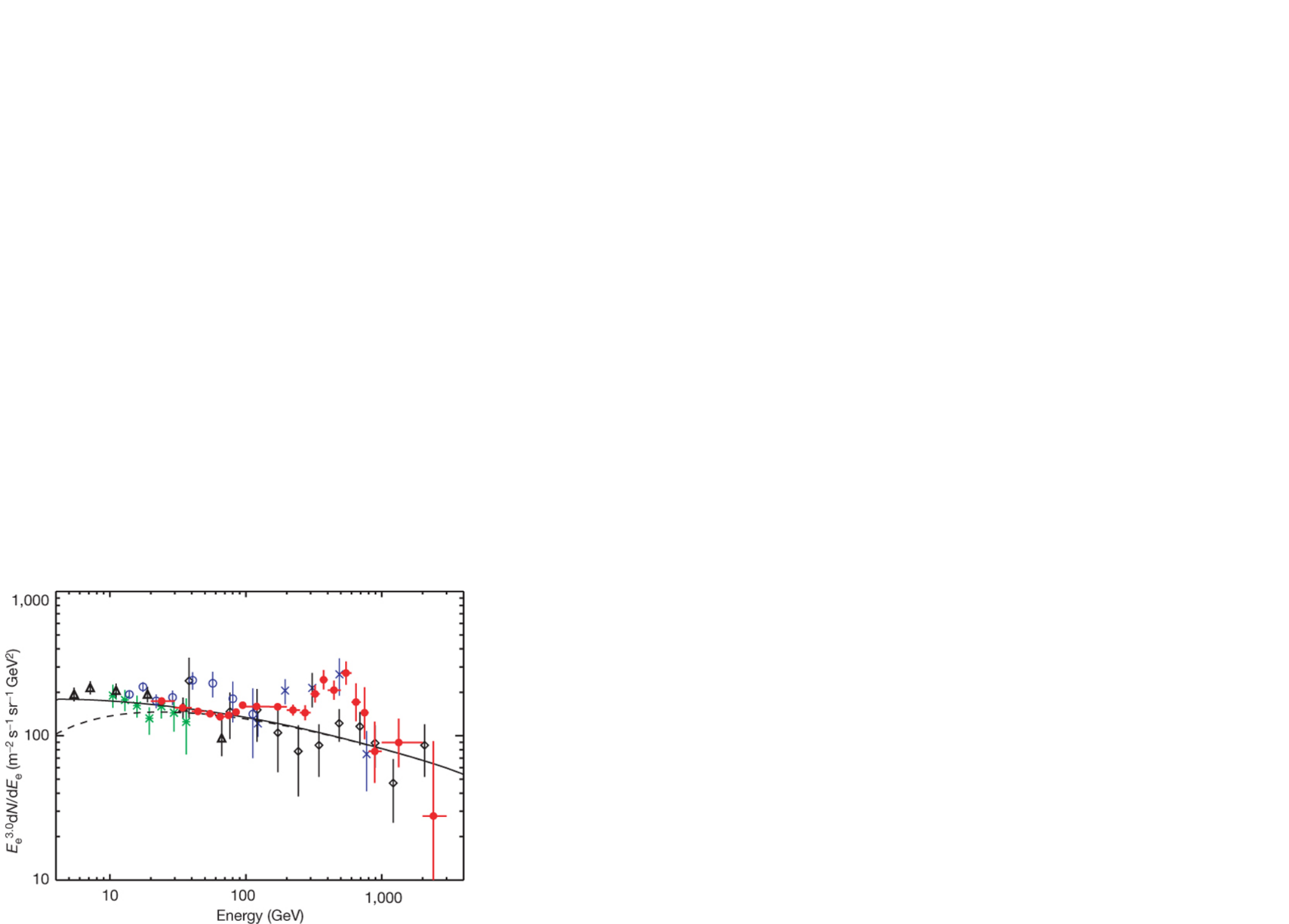}
\caption{{\bf Results from PAMELA and ATIC with theoretical models.}
The left panel shows PAMELA\cite{Adriani:2008zr}  positron fraction along with theoretical model. The solid black line shows a calculation by
Moskalenko \& Strong\cite{mos98} for pure secondary production of positrons during the propagation of cosmic-rays in the
Galaxy. The right panel shows the differential electron energy spectrum measured by ATIC\cite{:2008zzr} (red filled circles)
compared with other experiments and also with theoretical prediction using the GALPROP\cite{Strong:2001fu} code (solid line). The other data points are from
AMS\cite{Aguilar:2002ad}(green stars), HEAT\cite{Barwick:1997kh} (open black triangles), BETS\cite{Torii:2001aw} (open blue circles), PPB-BETS\cite{Torii:2008xu} (blue crosses) and emulsion chambers (black open diamonds) and the dashed curve at the beginning is the spectrum of solar modulated electron. All the data points have uncertainties of one standard deviation. The ATIC spectrum is scaled by $E_e^{3.0}$. The figures of PAMELA and ATIC are reproduced from their original papers cited above.
\label{pamela}}
\end{figure}

 The satellite-based \textit{Payload for Antimatter Matter Exploration and Light-nuclei Astrophysics} or
 PAMELA  collects cosmic ray protons, anti-protons, electrons, positrons and also light nuclei like
 Helium and anti-Helium.  One of the main strengths of PAMELA is that it could distinguish between
 electrons and anti-electrons, protons and anti-protons and measure their energies accurately.
 The sensitivity of the experiment in the positron channel is up to approximately 300 GeV and in the anti-proton
 channel up to approximately  200 GeV. Since it was launched in June 2006, it was placed in an elliptical orbit
 at an altitude ranging between $350 - 610$ km with an inclination of 70.0\textdegree.  About 500 days of data was
 analyzed and recently presented.  The present data is from 1.5 GeV to 100 GeV has been published in the
 journal Nature \cite{Adriani:2008zr}. In this paper, PAMELA reported an excess of positron flux compared to earlier experiments.
 In the left panel of the Fig.~\ref{pamela}, we see PAMELA results along with the other existing results. The
 y-axis is given by $\phi(e^+) / (\phi(e^-) + \phi(e^+) ) $, which $\phi$  represents the flux of the corresponding particle.
 According to the analysis presented by PAMELA, the results of PAMELA are consistent with the earlier experiments
 up to 20 GeV, taking  into consideration the solar modulations between the times of PAMELA and previous experiments.
 Particles with energies up to 20 GeV are strongly effected by solar wind activity which varies with the solar cycle.
 On the other hand, PAMELA has data from 10 GeV to 100 GeV, which sees an increase in the positron flux (Fig.~\ref{pamela}). The only
 other experimental data in this energy regime (up to 40 GeV) are the AMS and  HEAT, which while having large errors are consistent with 
 the excess seen by PAMELA. In the low energy regime most other experiments are in accordance with each other but have large error bars.

\begin{figure}[ht]
 \includegraphics[width=0.80\textwidth]{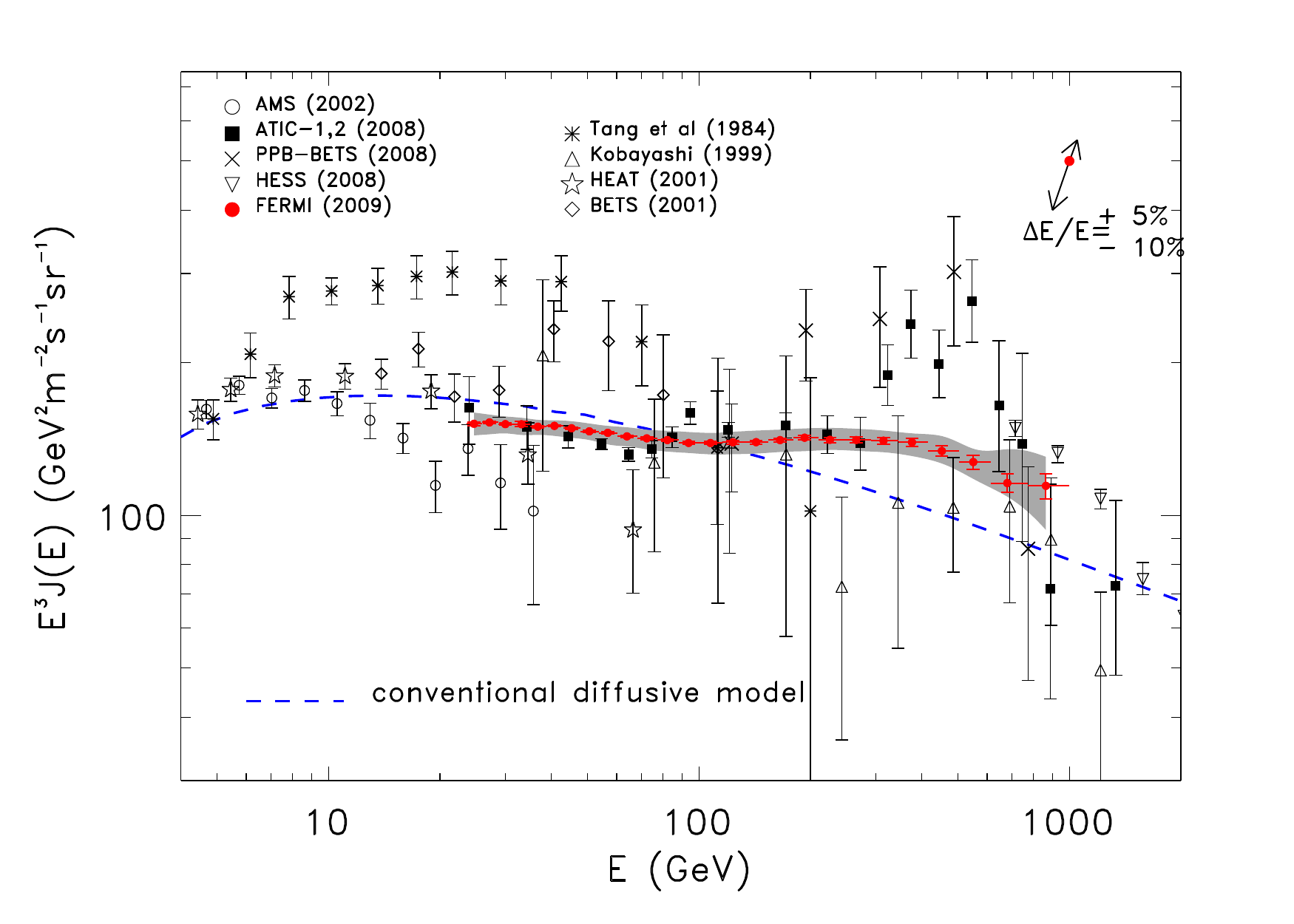}
\caption{{\bf The Fermi LAT CR electron spectrum.}
The red filled circles shows the data from Fermi along with the gray bands showing systematic errors. The dashed line
correspond to a theoretical model by Moskalenko \textit{et al.} \cite{Strong:2004de}. The figure of FERMI is reproduced from their original paper cited above.
\label{fermi}}
\end{figure}

Cosmic ray positrons at these energies are expected to be from secondary sources \textit{i.e.} as result of interactions of primary cosmic rays (mainly protons and electrons) with interstellar medium.  The flux of this secondary sources can be estimated by numerical simulations.  There are several numerical 
codes available to compute the secondary flux, the most popular publicly available codes being GALPROP \cite{galprop,galprop2} and CRPropa \cite{crpropa}. 
These codes compute the effects of interactions and energy loses during cosmic ray propagation within galactic medium taking also in to account the
galactic magnetic fields. GALPROP solves the differential equations of motion either using a 2D grid or a 3D grid while CRPropa does the same using
a 1D or 3D grids.  While GALPROP contains a detailed exponential model of the galactic magnetic fields, CRPropa implements only extragalactic
turbulent magnetic fields. In particular CRPropa is not optimised for convoluted galactic magnetic fields. For this reason, GALPROP is best suited 
for solving diffusion equations involving low energy (GeV-TeV) cosmic rays in galactic magnetic fields.  

The main input parameters of the GALPROP code are the primary cosmic ray injection spectra, the spatial distribution of cosmic ray sources, the size
of the propagation region, the spatial and momentum diffusion coefficients and their dependencies on particle rigidity. These inputs are mostly
fixed by observations, like the interstellar gas distribution is based on observations of neutral atomic and molecular gas, ionized gas ; cross sections
and energy fitting functions are build from Nuclear Data sheets (based on Las Almos Nuclear compilation of nuclear crosssections and modern nuclear codes) 
and other phenomenological estimates. Interstellar radiation fields and galactic magnetic fields are based on various models present in literature. The uncertainties
in these inputs would constitute the main uncertainties in the flux computation from GALPROP\footnote{Some codes   are constructed 
to fix the various parameters of their own cosmic ray propagation model. See for example, DRAGON \cite{dragon}.
Here one can fix the diffusion coefficients from  PAMELA and other experimental data.}.
Recently, a new code called CRT which emphasizes more on the minimization of the computation time was introduced. Here most of the input parameters are 
user defined \cite{crt}. Finally,  using the popular Monte Carlo routine GEANT \cite{geant} one can construct cosmic ray propagation code as has been
done by \cite{desorgher1,strumia}.  On the other hand, dark matter relic density calculators like DARKSUSY \cite{Gondolo:2004sc} also compute cosmic ray
propagation in the galaxies required for indirect searches of dark matter. It is further interfaced with GALPROP.  

 In summary,  GALPROP is most suited for the present purposes  \textit{i.e,}  understanding  of PAMELA and ATIC data which is mostly in the GeV-TeV range. 
 It has been shown the  results from these experiments  do not vary much if one instead chooses to use a GEANT simulation. In fact, 
 most of the experimental collaborations  use GALPROP for their predictions of secondary cosmic ray spectrum. 
 In the left panel of Fig.~\ref{pamela}, the expectations based on GALPROP are given as a solid line running across the figure. From the figure it is obvious that PAMELA
  results show that the positron fraction increases with energy compared to what GALPROP  expects. The excess in the positron fraction as measured by PAMELA with 
  respect to GALPROP indicates that this could be a result due to new primary sources rather than secondary sources \footnote{For an independent analysis which
  confirms the PAMELA excess,  please see, \cite{delahaye2}.}  This new primary source could be either dark 
  matter decay/annihilation or a nearby astrophysical object like a pulsar. Before going to the details of the interpretations, let us summarize the results from ATIC and FERMI also.

 \textit{Advanced Thin Ionization Calorimeter} or in short ATIC is a balloon-borne experiment to measure
 energy spectrum of individual cosmic ray elements within the region of GeV  up to almost a TeV (thousand GeV)
  with high precision. As mentioned,  this experiment was designed to be a high-resolution and high statistics experiment in
  this energy regime compared to the earlier ones. ATIC  measures all the  components of the
  cosmic rays such as  electrons,
  protons (and their anti-particles) with high energy resolution, while  distinguishing well between electrons
  and protons.  ATIC (right panel in Fig.~\ref{pamela}) presented its primary cosmic ray electron (\el + \ps) spectrum between the
  energies 3 GeV to about 2.5 TeV\footnote{The cosmic ray electrons follow a power law spectrum, the index being $\sim$ $-3.0$. Thus
  it is normalized by a factor $E^{3.0}$ .}.  The results show that the spectrum while agreeing with the  GALPROP expectations 
  up to 100 GeV,  show a sharp increase above 100 GeV.  The total flux increases till about 600 GeV where it peaks and then sharply falls 
  till about 800 GeV. Thus, ATIC sees an excess of the primary cosmic ray (\el + \ps) spectrum between the energy range $300-800$ GeV.
  The rest of the spectrum is consistent with the expectations within the errors.  What is interesting about such peaks
  in the spectrum is that,  if they are confirmed they could point towards a Breit-Wigner resonance in dark matter annihilation
  cross section with a life time as given by its width.  As we will discuss in the next section, this possibility is severely constrained
  by the data from the FERMI experiment. 

Another ground based experiment sensitive to cosmic rays within this energy range is H.E.S.S which can measure gamma rays from few hundred GeV to few TeV. 
This large reflecting array telescope operating from Namibia has presented data (shown in figure \ref{fermi})
 from $600$ GeV to about 5 TeV. It could confirm neither the \lq{}peaking\rq{} like  behavior at 600 GeV nor  the sharp cut-off at 800 GeV
  of the ATIC data.  The ATIC results can be made consistent with those of HESS. This would require
 a $15\%$ overall normalisation of the HESS data. Such a normalisation is well within the uncertainty of the energy resolution of HESS. 
  However notice that HESS data does not have a sharp fall about and after 800 GeV.

The Large Area Telescope (LAT) is one of the main components of the Fermi Gamma Ray Space Telescope, which was launched in June 2008. 
Due to its high resolution and high statistical capabilities, it has  been one of the most anticipated experiments in the recent times. 
Fermi can measure Gamma rays between 20 MeV and 300 GeV with high accuracy and primary cosmic ray electron (\el + \ps) flux between 20 GeV and 1 TeV. 
The energy resolution averaged over the LAT acceptance is 11\% FWHM (Full-Width-at-Half-Maximum) for 20-100 GeV, increasing to 13\% FWHM for 150-200 GeV. 
The photon angular resolution is less than 0.1\textdegree~over the energy range of interest (68\% containment). 
The FERMI-LAT collaboration has recently published its six month data on the primary cosmic ray electron flux.  
More than 4 million electron events above 20 GeV were selected in survey (sky scanning) mode from 4 August 2008 to 31 January 2009.
The systematic error on the absolute energy of the LAT was determined to be $^{-10\%}_{+5 \%}~$ for 20-300 GeV. 
Please see  Table I for more details on the errors in \cite{Abdo:2009zk}.
In Fig.~\ref{fermi} we reproduce the result produced by the FERMI collaboration. They find that the primary cosmic ray electron spectrum more or less goes along the expected lines up to 100 GeV (its slightly below the expected flux between 10 and 50 GeV), however above 100 GeV, there is strong signal for an excess of the flux ranging up to 1 TeV. The FERMI data thus confirms the excess in the electron spectrum which was seen by ATIC, the excess however has a much flatter profile with respect to the peak seen by ATIC.  Thus, ATIC could in principle signify a \lq{}resonance\rq{} in the spectrum, whereas FERMI cannot. However, in comparing both the spectra from the figures presented above, one should keep in mind that the FERMI excess is in the total electron spectrum ($e^+ + e^-$ ) whereas the ATIC data is presented in terms of positron excess only. If the excess in FERMI is caused by the excess only through excess positrons, one should expect that the FERMI spectra to also have similar \lq{}peak\rq{} like behavior at 600 GeV. From Fig.(\ref{fermi}), where both FERMI and ATIC data are presented, we see that the ATIC data points are far above that of FERMI's. 

\section{The Interpretations  }
Lets now summarise the experimental observations \cite{strumia} which would require an interpretation :
\begin{itemize}
\item The excess in the flux of positron fraction $\left(\tfrac{\phi(e^+)}{\phi(e^-) + \phi(e^+)}\right)$    measured by PAMELA up to 100 GeV.
\item The lack of excess in the anti-proton fraction measured by PAMELA up to 100 GeV.
\item The excess in the total flux $\left(\phi(e^-) + \phi(e^+)\right)$ in the spectrum above 100 GeV seen by FERMI, HESS {\it etc.} While below 100 GeV, the measurements have been consistent with GALPROP expectations.
\item The absence of \lq{}peaking\rq{} like behavior as seen by ATIC, which indicates a long lived particle, in the total electron spectrum measured by FERMI.
\end{itemize}

Two main interpretations  have been put forward:
(a) A nearby astrophysical source which has a mechanism to accelerate particles to high energies and (b)
A dark matter particle which decays or annihilates leading to excess of electron and positron flux.
Which of the interpretations is valid will be known within the coming years with enhanced data from both PAMELA and FERMI.  Let us now turn to both the interpretations:

Pulsars and supernova shocks have been proposed as likely astrophysical local sources of energetic particles that could  explain the observed excess of the positron fraction \cite{Hooper:2008kg, Blasi:2009hv}. In the high magnetic fields present in the pulsar magnetosphere, electrons can be accelerated and induce an electromagnetic cascade through the emission of curvature radiation\footnote{The curvature radiation arises due to relativistic, charged particles moving around curved magnetic field lines, see for details Gil {\it et al.}\cite{Gil:2003ug}.}. This can lead to a production of high energy photons above the threshold for pair production; and on combining with  the number density of pulsars in the Galaxy,  the resulting emission can explain the observed positron excess \cite {Hooper:2008kg}. The energy of the positrons tell us about  the site of their origin and their propagation history \cite{coutu99}. The cosmic ray positrons  above 1 TeV could be primary and arise due to a source like a young plusar within a distance of 100 pc \cite{ato95}. This would also naturally explain the observed anisotropy, as argued for two of the nearest pulsars, namely $B0656+14$ and the $Geminga$  \cite{bue08,Yuksel:2008rf}. On a similar note,  diffusive shocks as in a supernova remnant  hardens the spectrum,  hence this process can explain the observed positron excess above 10 GeV as seen from PAMELA \cite{Ahlers09}. 

Another possible astrophysical source that has been proposed is
the pion production during acceleration of hadronic cosmic rays
in the local sources \cite{Blasi:2009hv}. It has been argued  \cite{Mertsch:2009ph} that the
measurement of secondary nuclei produced by cosmic ray spallation can confirm whether this process or pulsars
are more important as the production mechanism.
It has been  show that the present data from ATIC-II supports the hadronic model and can account for the entire positron excess
observed.

If the excess observed by PAMELA, HESS  and FERMI is not  due to some  yet
not fully-understood  astrophysics but is a signature of the dark matter, then there
are two main processes through which such an excess can occur:

\begin{enumerate}
 \item [(i)] The annihilation of dark   matter particles into Standard Model (SM) particles and
 \item[(ii)] The  decay  of the dark matter particle  into  SM particles.
\end{enumerate}

Interpretation in terms of annihilating dark matter, however, leads to conflicts with cosmology. The 
observed excesses in the PAMELA/FERMI data would set a limit on the product of annihilation 
cross section and the velocity of the dark matter particle in the galaxy (for a known dark matter
density profile). Annihilation of the dark matter particles also happens in the early universe with
the same cross section but at much larger velocities for particles (about 1000 times the particle velocities in galaxies).
The resultant relic density is not compatible with observations. The factor $\sim 1000$ difference in the 
velocities should some how be compensated in the cross sections. This can be compensated by considering 
``boost" factors for the particles in the galaxy which can enhance the cross section by several orders of
magnitude. The boost factors essentially emanate from assuming local substructures for the dark 
matter particles, like clumps of dark matter and are typically free parameters 
of the model  (see however, \cite{delahaye1}).  Another mechanism which goes by the name
Sommerfeld mechanism can also enhance the annihilation cross sections. 
For very heavy dark matter ( with masses  much greater than the relevant gauge boson masses) 
 trapped in the galactic potential,  non-perturbative effects could push the annihilation 
 cross-sections to much larger values.  For SU(2) charged dark matter, the masses of dark
 matter particles should be $\gg M_W$ \cite{hisano}.  The Sommerfeld mechanism is more 
 general and  applicable to other (new) interactions also\cite{hannestead}. 
 Another way of avoiding conflict with cosmology would be
to consider non-thermal production of dark matter in the early universe\footnote{Non-thermal production 
typically refers to production mechanisms through decays of very heavy particles like inflaton. 
See for example \cite{riotto1}.}. Before the release of FERMI data, the annihilating dark matter model with
a  very heavy dark matter  $\sim \mathcal{O}(2-3)\ $TeV  was  much in favour to explain  the
``resonance peak" of the ATIC and  the excess in PAMELA data. Post FERMI, whose data does not
have sharp raise and fall associated with a resonance, the annihilating dark matter interpretation
has been rendered incompatible. However,  considering possible variations in the local astro physical
background profile  due to presence of local cosmic ray accelerator, it  has  been shown that it is 
still possible to explain the observed excess, along with FERMI data  with annihilating dark matter. 
The typical mass of the dark matter particle could lie even within  sub-TeV 
region \cite{Dodelson:2009ih,Belikov:2009cx,Cholis:2009gv,Hooper:2009cs} and as low as $30-40$ GeV
\cite{Goodenough:2009gk}. Some more detailed analysis can be found in \cite{Pato:2009fn}.

Several existing BSM physics models of annihilating dark matter become highly constrained or ruled out if one requires to explain PAMELA/ATIC 
and FERMI data. The popular supersymmetric DM candidate neutralino with its annihilating partners such as  chargino, stop, stau {\it etc.}, can explain the cosmological relic density but not the excess observed by PAMELA/ATIC.  Novel models involving a new \textit{\lq{}dark force\rq{}}, with a gauge boson having mass of about 1 GeV \cite{ArkaniHamed:2008qn}, which predominantly decays to \textit{leptons}, together with the so-called Sommerfeld enhancement  
seem to  fit the data well. The above class of
models, which are extensions of standard model with an additional $U(1)$ gauge group, caught the imagination of the theorists
\cite{Katz:2009qq,Cholis:2008vb,Cholis:2008hb,Cholis:2008qq}.  A similar supersymmetric version of this mechanism
where the neutralinos in the MSSM can annihilate to a scalar particle, which can then decay the observed excess in the cosmic
ray data \cite{Hooper:2009gm}. Models involving Type II seesaw mechanism \cite{Gogoladze:2009gi} have  also been  considered recently where neutrino mass generation is linked with the positron excess. In addition to the above it has been shown that extra dimensional models with KK gravitions can also produce the excess \cite{Hooper:2009fj}\footnote{Some of the first simulations using PYTHIA and DARK SUSY for the KK gravition has been done in \cite{hooperkribs}. Similar study for SUSY can be found in \cite{susy01}. These have been done when HEAT results have shown an excess though in a less statistically significant way.}. Models with  Nambu-Goldstone bosons as  dark matter have been studied in \cite{murayama}.

In the case of decaying dark matter, the relic density constraint of the early universe is not
applicable, however, the lifetime of the dark matter particle (typically of a mass
of $ \mathcal{O}(1)$ TeV) should be much much larger than ($\sim 10^{9}$ times)
 the age of the universe\cite{strumia}.  Such a particle can fit the data well.
 A crucial difference  in this picture with respect to the annihilation picture is that the
 decay rate is directly  proportional to the density of the dark matter ($\rho$), whereas
 the annihilation rate is proportional to  its square, ($\rho^2$). The most promising
candidates in the decaying dark matter seem to be a  fermion (scalar) particle
decaying in to $W^\pm l^\pm $ {\it etc.} ($W^+ W^-$ {\it etc.}) \cite{Ibarra:2009dr,Ibarra:2009nw,Ibarra:2008jk,Nardi:2008ix}.
 In terms of the BSM physics, supersymmetric models with a heavy gravitino and small R-parity violation have been proposed
 as candidates for decaying dark matter \cite{Buchmuller:2007ui}.  A heavy neutralino with $R$-parity violation can
also play a similar role \cite{Gogoladze:2009kv} stated above. A recent more general model independent analysis has shown
that, assuming the GALPROP background, gravitino decays cannot simultaneously explain both PAMELA and FERMI excess. However,
the presence of additional astrophysical sources can change the situation \cite{Buchmuller:2009xv}. Independent of the
gravitino model, it has been pointed out that, the decays of the Dark Matter particle could be new signals for unification
where the Dark Matter candidate decays through dimension six operators suppressed by two powers of GUT scale
\cite{Arvanitaki:2008hq, Arvanitaki:2009yb,Buckley:2009kw}. Finally,  there has also been some discussion about the possibilities 
of dark matter consisting of not one particle but two particles, of which one is the decaying partner. 
This goes under the name of \lq{}{\it two-component dark matter}\rq{} and analysis of this scenario has been presented by  \cite{Fairbairn:2008fb}.  

We have so far mentioned just a sample of the theoretical ideas proposed in the literature. Several other equally interesting 
and exciting ideas have been put forward, which have not been presented to avoid the article becoming too expansive. 

\section{Outlook and Remarks}

An interesting aspect about the present situation is that, future data from PAMELA and
FERMI could distinguish whether the astrophysical interpretation \textit{i.e.} in terms
of pulsars or the particle physics interpretation in terms of dark matter is valid \cite{Malyshev:2009tw}. PAMELA
is sensitive to up to 300 GeV in its positron fraction and this together with the measurement
of the total electron spectrum can strongly effect the dark matter interpretations. FERMI with its
improved statistics, can  on the other hand look for anisotropies within its data \cite{Grasso:2009ma}
which can exist if the pulsars are the origin of this excess. Further measurements of the anti-Deuteron could possibly gives
us a hint why there is no excess in the anti-Proton channel \cite{Kadastik:2009ts}.  Similarly neutrino physics
experiments could give us valuable information on the possible models\cite{hisano2}. Finally,  the 
Large Hadron collider could also give strong hints on the nature of dark matter through direct
production \cite{LHCdm}.

As we have been preparing this note, there has been news from one the experiments called
CDMS-II (Cryogenic Dark Matter Search Experiment)\cite{cdms2}. As mentioned before
this experiment conducts direct searches for WIMP dark matter by looking at collisions of
WIMPs on super-cooled nuclear target material.  The present and final analysis of this
experiment have shown two events in the signal region, with the probability of observing
two or more background events in that region  being close to $23\%$. Thus, while these results 
are positive and encouraging, they are not conclusive. However these results already set 
a stringent upper bound on the WIMP-nucleus cross section for a WIMP mass of around
70 GeV.  The exclusion plots in the parameter space of WIMP cross section and WIMP mass
are presented in the paper \cite{cdms2}. The interpretations of this positive signal are quite different
compared to the signal of PAMELA and FERMI. While PAMELA and FERMI as we have seen
would require severe modifications for the existing beyond standard model (BSM) models of Dark Matter, CDMS results
if confirmed would prefer the existing BSM dark matter candidates like neutralino of the supersymmetry.
There are ways of making both PAMELA/FERMI and CDMS-II consistent through dark matter
interpretations, however, we will not discuss it further here.  Finally, it has been shown that  it is possible
to make CDMS-II results consistent with DAMA annual modulation results by assuming a spin-dependent
inelastic scattering of WIMP on Nuclei \cite{koppzupan}. 

In the present note, we have tried to convey  exciting developments
which have been happening recently within the interface of astrophysics and particle physics,
especially on the one of the most intriguing subjects of our time, namely, the \textit{Dark Matter}. 
Though it has been proposed about sixty years ago, so far we have not have any conclusive 
evidence of its existence other than through gravitational interactions, or we do not of its fundamental
composition. Experimental searches which have been going on for decades have not bore fruit 
in answering either of these questions. For these reasons, the present indications from PAMELA
and FERMI have presented us with a unique opportunity of unraveling at least some of mystery
surrounding the dark matter.  These experimental results, if they hold and get confirmed as due
to dark matter, would strongly modify  the way dark matter was perceived in the scientific community. 
As a closing point, let us note that there are several new experiments being planned to explore the
dark matter either directly or indirectly and thus some information about the nature of the
dark matter  might just around the corner. 


\acknowledgements{}
We thank PAMELA collaboration, ATIC collaboration and FERMI-LAT collaboration for giving us
permission to reproduce their figures. We thank Diptiman Sen for a careful reading of this article
and useful comments. C. J. would like to thank Gary Mamon for illuminating discussions regarding
 the search for dark matter in elliptical galaxies and clusters. We thank A. Iyer for bringing to our
 notice a reference. Finally, we thank the anonymous referee for suggestions and comments which
have contributed in improving the article.



\end{document}